\documentclass[twocolumn,showpacs,preprintnumbers,amsmath,amssymb]{revtex4}

\usepackage{graphicx}
\usepackage{dcolumn}
\usepackage{bm}

\begin{document}

\title{Defect controlled vortex generation in current-carrying narrow
superconducting strips}

\author{D.Yu. Vodolazov$^{1,2}$}
\affiliation{$^1$ Institute for Physics of Microstructures,
Russian Academy of Sciences, 603950,
Nizhny Novgorod, GSP-105, Russia \\
$^2$ Lobachevsky State University of Nizhny Novgorod, 23 Gagarin
Avenue, 603950 Nizhny Novgorod, Russia }
\author{K. Ilin$^{3}$, M. Merker$^{3}$, M. Siegel$^{3}$}
\affiliation{$^3$ Institute of Micro- and Nanoelectronic Systems
(IMS), Karlsruhe Institute of Technology (KIT), Hertzstr. 16,
76187 Karlsruhe, Germany}

\date{\today}


\begin{abstract}
We experimentally study effect of single circular hole on the
critical current $I_c$ of narrow superconducting strip with width
$W$ much smaller than Pearl penetration depth $\Lambda$. We found
nonmonotonous dependence of $I_c$ on the location of a hole across
the strip and a weak dependence of $I_c$ on radius of hole has
been found in case of hole with $\xi \ll R \ll W$ ($\xi$ is a
superconducting coherence length) which is placed in the center of
strip. The observed effects are caused by competition of two
mechanisms of destruction of superconductivity - the entrance of
vortex via edge of the strip and the nucleation of the
vortex-antivortex pair near the hole. The mechanisms are clearly
distinguishable by difference in dependence of $I_c$ on weak
magnetic field.
\end{abstract}

\maketitle

\section{Introduction}

The maximal current, which can pass through the superconducting
strip without dissipation, is equal to $I_{dep} = j_{dep}Wd$,
where $W$ is a width and $d$ is a thickness of the strip and
$j_{dep}$ is the depairing current density. However in
experiments, this value can be reached only in relatively thin ($d
<\lambda$, $\lambda$ is the magnetic field penetration depth) and
narrow strips with $W < \Lambda=2\lambda^2/d$ in which a current
density is distributed uniformly across the width and thickness
\cite{Likharev}. In opposite situation of wide and thick strips
($d>\lambda$, $W \gg \Lambda$), an expelling of magnetic field,
which is induced by an applied current, from the interior of the
strip leads to piling up of current density at the edge of strip
and the maximal critical current becomes smaller than $I_{dep}$.
Moreover, even in the narrow and thin superconducting strips with
uniformly distributed current density the measured critical
current $I_c$ is smaller than $I_{dep}$
\cite{JETP_Andr,PRB_Rusanov,Nawaz}. The suppression of
experimentally measured critical current in such structures can be
explained by a presence of defects in the strip. Under defect we
understand a local perturbation of properties of the strip which
provides a local suppression (complete or partial) of the
superconducting order parameter $\Delta$ in the current-carrying
state of the strip. This local suppression of $\Delta$ can be due
to local variation of the thickness/width of the strip, its
material parameters such as the electron mean path and/or critical
temperature or such suppression of $\Delta$ may appear temporally
in the place of absorption of high energetic photons or particles.
The superconducting current avoids the place with suppressed
$\Delta$ and this phenomenon leads to nonuniform current
distribution, with the maximal current density near the defect
place. Now this phenomenon is known as a current crowding effect,
which was studied in details theoretically in \cite{C&B} and
confirmed experimentally in \cite{Henrich,Hort,Adami,Akh} on model
systems of narrow strips with sharp bends which play a role of
defects.

In the recent theoretical work \cite{SUST_DV}, the effect of a
single defect, which is located close or far from the edge of the
superconducting strip with finite width, on generation of
resistive state was studied in frame of Ginzburg-Landau approach.
The defect was modelled by the circle with locally suppressed
superconductivity - we call it later as a 'hole'. It has been
shown that the mechanism of destruction of superconducting state
by applied current in a narrow strip depends on position of the
hole with respect to the edge of strip. When the hole sits close
to the edge, the self-generated vortices enter the hole via the
edge of the film and then they move across the superconductor. In
this sense it resembles effect of edge defect
\cite{Buzdin,Melnikov,Vodolazov_def,Koshelev}. When the hole is
located pretty far from the edges, the vortex-antivortex pair is
nucleated inside the hole and a motion of the pair destroys the
superconducting state \cite{JETP_1983,SUST_DV}. Domination of the
first or the second mechanism of destruction of superconductivity
leads to extremely nonmonotonic dependence of critical current
$I_c$ on the position of the hole (see Fig. 4 in Ref.
\cite{SUST_DV}).

The interesting effects arise when the hole is located in the
center of strip. First of all the theory predicts that in a
certain range of the hole's radius $\xi \ll R \ll  W$ the critical
current is independent of $R$ and equals to about $0.5 I_{dep}$
\cite{C&B,PRB_DV_Z}. This counterintuitive result (one could
expect that $I_c = I_{dep}(1-2R/W)$ in case of uniform current
distribution near the hole and $I_c \simeq I_{dep}$ when $R\ll W$)
is the direct consequence of the current crowding near the
circular hole. Secondly, in this particular situation the small
magnetic field $B$ weakly affects the critical current in
comparison to a strip without hole. This weakening of the $I_c(B)$
dependence is because the resistive state is determined by the
vortex-antivortex nucleation near the hole and not at the edges of
a strip where Meissner currents are maximal. Moreover, due to
opposite direction of the generated Meissner currents at the
opposite edges of strip, the influence of external magnetic field
on $I_c$ of strips is dependent on location of the hole. This
effect leads to increase of $I_c$ when the hole is near to one of
the edges and to decrease of $I_c$ when the hole is at the
opposite edge (see Fig. 9 in \cite{SUST_DV}).

We stress here the difference between the effect of defects at
relatively large magnetic field (when there is dense vortex matter
in the strip) and at zero or small magnetic fields $B\lesssim B_s
\simeq \Phi_0/2\pi \xi W $(when there is no field induced
vortices) in current-carrying superconductors. In the former case,
the vortices, which are created by the externally applied magnetic
field, can be effectively pinned by defects and, thereby, this
strengthens transport properties of superconductor. Pinning by
intrinsic and artificially created defects has been extensively
studied in numerous theoretical and experimental works
\cite{Reichhardt,Castellanos,Velez}. In the case of small or even
zero magnetic field, the superconductivity is weakened near the
defect due to current crowding. When the applied current exceeds
some critical value the vortices are nucleated near this weak
place and their motion destroys the superconducting state
\cite{JETP_1983,SUST_DV}.

Comparison of the theoretical predictions \cite{SUST_DV} with
experimental results, which are obtained in real structures, is
complicated. It is very difficult to find a single natural
(intrinsic) defect in a film which localizes far enough from a lot
of other similar defects and then arrange experiment to study
transport properties which are determined by this particular
defect and are not smeared out by collective effect of other
defects. In case of local suppression of superconductivity caused
by absorbed photon, deterministic localization of absorption site
is not possible at all due to probabilistic nature of the photon
absorption.

In the present work we made holes of different size and placed
them in different places across the superconducting strip. Such a
model system mimics the main effect of the real defect - current
crowding effect and it allows us in controllable manner to study
the role of the defects on the critical current and mechanisms of
destruction of the superconducting state.

We have experimentally found the non-monotonous dependence of
$I_c$ of strips with the hole with constant radius $R$ on the
position of this hole with respect to the edge of the strip. We
have also shown that the dependence of $I_c$ on weak magnetic
fields in case of mechanism of vortex-antivortex nucleation in
vicinity of the hole is weaker than the $I_c(B)$-dependence which
is determined by a single-vortex penetration through the edge of
superconducting strip in agreement with theoretical predictions
\cite{SUST_DV}. Our results allow not only to clarify the role of
defects on suppression of the critical current at zero and low
magnetic fields ($B\lesssim B_s$, Meissner state) but they also
help to understand the mechanism of photon detection by
current-carrying superconducting strips.

\section{Experiment}

NbN films with thickness about 5 nm were deposited on a heated
sapphire substrate by DC reactive magnetron sputtering of pure Nb
target in a gas mixture atmosphere of argon and nitrogen. For the
used deposition conditions of the total pressure 2.8 $10^{-3}$
mbar, partial pressure of nitrogen 3.2 $10^{-4}$ mbar and sputter
current 150 mA, films with the critical temperature $T_c \simeq$
13 K, square resistance of about 300 Ohm  and residual resistance
ratio RRR = 0.89 were deposited.

Patterning of films has been done by electron-beam lithography and
ion-milling technique. The films were patterned into single-bridge
structures with a width about 1 $\mu$m and a length 20 $\mu$m
which were embedded between two millimeter sized contact pads. The
gradual transition between the bridge and the contacts with radius
of curvature $r$ = 5 $\mu$m was realized to avoid a current
crowding at T-shape connections which were considered in details
in \cite{C&B}.
\begin{figure}[hbtp]
\includegraphics[width=0.44\textwidth]{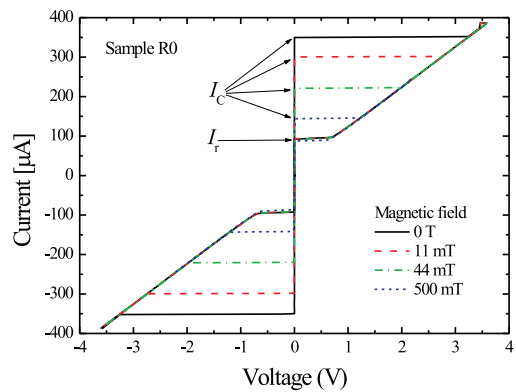}
\caption{Current-voltage characteristics of the superconducting
strip without hole at different magnetic fields indicated in the
legend. The arrows indicate correspondent critical currents $I_c$
and current $I_r$ of recovery of the bridge into superconducting
state.}
\end{figure}

Two series of samples (24 bridges each) have been fabricated on
separated 10 mm squared chips. In the first series of samples,
circle with radius $R \simeq$130 nm was etched inside a bridge.
The etched circles were always placed at a half-length of bridge.
The coordinate of circles across the width of bridge was varied
with respect to the left edge of bridge with a step  $\delta x
\simeq$ 35 nm between two subsequent positions (see the insets in
Fig. 2). Through the paper samples which belong to this series of
bridges will be named XNN where NN is a distance in nanometer
between left edge of a strip and center of a hole which is etched
inside the strip. The radius of circles with a center placed at
the edge of bridges (in this case it is better to talk about
semi-circles, see the top left inset in Fig. 2) was by a factor
$\sqrt{2}$ larger i.e. about 185 nm. This was done to keep
constant the area with suppressed order parameter independently of
location of the hole.

In the second series of samples, the position of etched circle
(the hole) was kept constant in the middle (half-length,
half-width) of a strip but the radius of hole was varied by an
order of magnitude from about 29 nm up to 225 nm. In the paper,
samples, which belong to this series, will be named RNN where NN
is a radius of etched hole in nanometer. Real dimensions of
fabricated structures (width of bridge, $W$, radius of hole, $R$,
and its position with respect to the left edge of bridge,
coordinate $x$) were measured after patterning by a scanning
electron microscopy (SEM).

All bridges were characterized by their critical temperature of
superconducting transition. The $T_c$ value was determined as a
temperature at which resistance of measured structure drops below
0.1$\%$ of the normal state resistance measured at temperatures
approximately twice higher than $T_c$. The obtained spread of
$T_c$ among bridges in each series was about 0.3 K which can be
attributed to slight variation of superconducting strength over
the film caused by, for example, non-uniformity of thickness of
deposited film. In case of ultra-thin films with thickness about
the coherence length which is in case of NbN is about 4-5 nm
\cite{PRB14Ilin}, even a small difference in thickness leads to
relatively large change in $T_c$ \cite{PRB09Sem, PRB13Noat}.

The current-voltage characteristics (IV-curves) were measured in a
cryogenic-free system at $T= $5.1 K in a DC current-bias mode. The
magnetic field up to 1 T generating by a superconducting solenoid
was applied normally to a sample surface. From the previous
studies \cite{PRB14Ilin} we know that the field, which is required
for penetration of magnetic vortex into a micrometer wide bridge
made from thin NbN film, is in the range of few tens millitesla
and therefore the available in our experiments range of $B$ is
enough for detailed investigations of bridges in Meissner and in
the vortex states. The minimum step of variation of magnetic field
was 1 mT which is determined by accuracy and stability of a used
power supply of superconducting solenoid.
\begin{figure}[hbtp]
\includegraphics[width=0.52\textwidth]{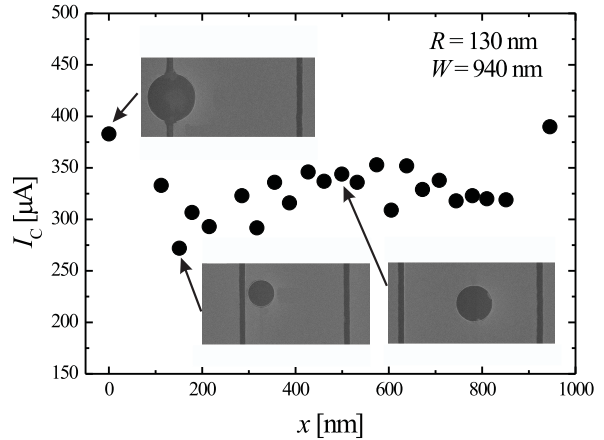}
\caption{Experimental dependence of the critical current on the
position of the hole in the strip (series "X"). Width of the strip
is 940 nm, radius of the hole is 130 nm $\sim W/7$ ($R$=185 nm
when semi-hole is placed at the edge). In the insets we present
SEM images of the strips with the holes in different positions.}
\end{figure}

Typical IV-curves for four different magnetic fields are shown in
Fig.1. The curves were hysteretic independently of the magnetic
fields in the whole range of applied fields $B$. At a critical
current $I_c$, the bridge is characterized by a sharp jump from
superconducting to normal state. Recovery of superconductivity in
the bridge happens at current $I_r$ which is almost independent of
magnetic field as it is clearly seen in Fig. 1. Contrary, $I_c$
demonstrates significant decrease with increase of the applied
field. The accuracy of $I_c$  determination in our experimental
setup is 0.5 $\%$ of $I_c$, i.e. about 1 $\mu$A for studied range
of critical currents.
\begin{figure}[hbtp]
\includegraphics[width=0.52\textwidth]{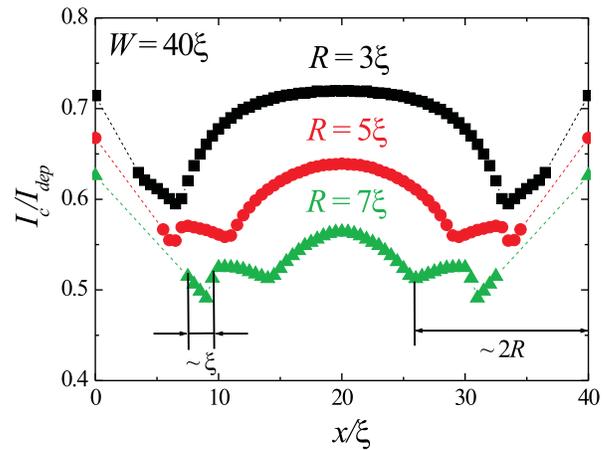}
\caption{Theoretical dependence of the critical current on
coordinate of hole's center in the strip with width $W=40 \xi$.
Results are shown for holes with three radii.}
\end{figure}

\section{Results}

The experimentally obtained dependence of the critical current
$I_c$ on position of the hole is shown in Fig. 2. It is seen that
$I_c$ is maximal when the hole is in the middle of the strip and
when the hole (semi-circle) is placed at the edge of the strip.
Shift of position of the hole from the center of the strip leads
to gradual decrease of $I_c$ which reaches the minimum values when
the hole is close to the edge. This experimental result
qualitatively coincides with the prediction of Ref.
\cite{SUST_DV}. In Fig. 3 we present results of theoretical
calculations which were performed in the framework of
Ginzburg-Landau model for different holes and strip with $W=40
\xi$ (details of numerical calculations are present in Ref.
\cite{SUST_DV}). In contrast to Ref. \cite{SUST_DV} we determine
the critical current as the current at which permanent vortex flow
starts in the strip with a hole. In Ref. \cite{SUST_DV} $I_c$ was
defined as the current at which the first vortex enters the hole
when it is located near the edge of the strip - due to assumption
that the 'hole' is created by the absorbed photon and it should
dissolves after some finite time interval. New definition leads
only to quantitative difference with result of Ref.
\cite{SUST_DV}: i) the local minimum in dependence $I_c(x)$ now
reaches when the hole sits on the finite distance from the edge;
ii) additional minima appear when the hole practically touches the
edge. These minima appear for relatively wide strips $W \gtrsim 30
\xi$ and large radiuses of the holes $R\gtrsim 3\xi$. The width of
the additional minima is about the coherence length (see Fig. 3)
and it is rather difficult to resolve it in the experiment due to
small $\xi \simeq 5$ nm in NbN. Small width of this minima ($\sim
\xi$) points on its origin - it appears due to suppression of the
order parameter in the narrow sidewalk between edge of the hole
and edge of the strip when its width becomes about $\xi$ (because
we model the hole as a region with locally suppressed
superconductivity). It results in decrease of the
'superconducting' width of the strip.

We also studied how $I_c$ depends on the radius of the hole when
it is placed in the center of the strip. Theoretical results are
present in Fig. 4 while experimental $I_c(R)$-dependence is shown
in Fig. 5. The theory predicts, that when $\xi \ll R \ll W$ the
critical current very weakly depends on $R$ (see inset in Fig. 4).
Calculations in the London model \cite{C&B,PRB_DV_Z} predict that
in this case $I_c/I_{dep}=1/2$ while our numerical results give a
little larger value $I_c/I_{dep} \simeq 0.65-0.7$. The difference
between theoretical results is most probably connected with the
characteristics of the hole. In Refs. \cite{C&B,PRB_DV_Z} the hole
was considered as a 'well' with jump of $\Delta$ at the edge while
in our model, due to proximity effect, $\Delta$ changes from zero
to its maximal value on the distance $\sim \xi$ near the hole's
edge.

In the experiment we find small variation of $I_c \simeq 284-335$
$\mu$A ($I_c/I_{dep} \simeq 0.42-0.49$) for holes with radius $R
\simeq 26-76$ nm ($R \simeq 5-15 \xi$) - see Fig. 5. The depairing
current $I_{dep}=679$ $\mu$A has been calculated in frame of
Ginzburg-Landau theory with account of the temperature dependent
correction factor which was obtained by Kupriyanov and Lukichev
\cite{Kupriyanov} for dirty limit superconductors (see Eqs. 3 and
4 in \cite{PRB14Ilin}). We have to note, that even for strips
without hole $I_c/I_{dep}\simeq 0.5$ (see Fig. 5) due to presence
of the intrinsic defects. These defects may be present also at the
edge of the hole which additionally suppresses the critical
current and it could be a reason for dispersion of $I_c$ for holes
with radius $R=26-76$ nm and smaller value of $I_c$ than expected
from the theory.
\begin{figure}[hbtp]
\includegraphics[width=0.48\textwidth]{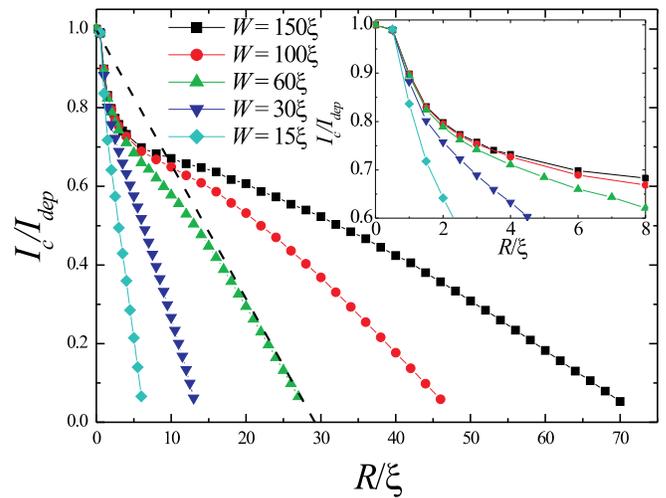}
\caption{Calculated dependence of $I_c$ on radius of hole placed
in the center of the superconducting strip with different width
indicated in the legend. Dashed line corresponds to linear
dependence $I_c/I_{dep}=1-2R/W$ for the strip with $W=60 \xi$. In
the inset we zoom the region of small radii $R\ll W$.}
\end{figure}

\begin{figure}[hbtp]
\includegraphics[width=0.48\textwidth]{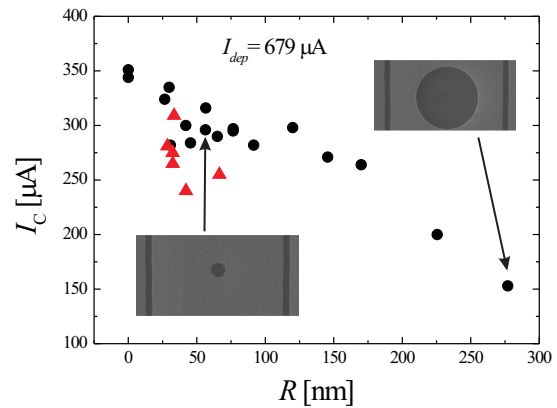}
\caption{The experimentally measured critical current of strips
($W$ = 980 nm) with hole of different radius, which is placed in
the center of the superconductor (series "R"). In the inset we
show SEM images of a center part of strip with hole. Triangles
mark the samples with linear $I_c(B)$ dependence at low $B$ which
indicates that vortices in these samples are generated at the
edges.}
\end{figure}

Theory \cite{SUST_DV} predicts, that when the hole is placed in
the center of the strip the resistive state starts via nucleation
of the vortex-antivortex pair near the hole. Direct experimental
visualization of this process is practically impossible at the
present time, due to very short time scale ($\sim$ picoseconds) of
vortex nucleation. However it could be checked indirectly, using
external magnetic field $B$. The applied magnetic field induces
screening current which is maximal near the edges and it is equal
to zero in the center of the strip. Therefore if resistive state
is realized via vortex entrance via edge of the film then applied
magnetic field strongly influences $I_c$. Contrary, if resistive
state is realized via generation of the vortex-antivortex pairs
near the central part of the strip the same magnetic field weakly
affects $I_c$. Experimental $I_c(B)$-dependencies will differ
significantly in dependence on location of penetration of vortex
or vortex-antivortex pair and, thereby, can be used for
determination of position of vortex generation in the strip at
$I>I_c$.

In Fig. 6 we show experimental dependencies $I_c(B)$ for strips
with and without hole in the center. For samples R26, R91 and R170
($R$ = 26, 91, 170 nm) there is plateau at low $B$ while for strip
without hole (sample R0) $I_c$ decays almost linearly at low $B$.
Therefore we conclude, that in the sample without the hole the
vortices are generated at the edge of the strip while in the
samples with the hole the vortex-antivortex pairs are generated
near the hole at $|B| \lesssim $ 5-18 mT (the larger the hole the
larger the threshold magnetic field). Magnetic fields larger the
threshold produce large screening currents at the edges of the
strip which, together with transport current, exceeds the current
density near the hole and the vortices start to be generated at
the edge. As a result the strong field dependence of $I_c$ is
recovered.
\begin{figure}[hbtp]
\includegraphics[width=0.53\textwidth]{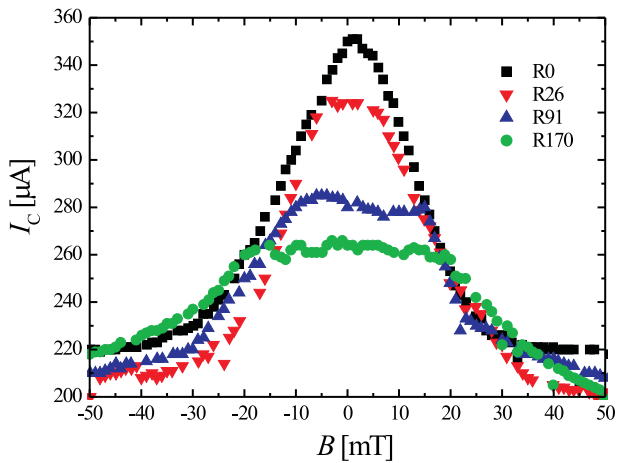}
\caption{The dependence of $I_c$ on magnetic field $B$ for strips
with hole of radii $R$ = 26, 91, 170  nm (R26, R91, R170) in the
center of the superconductor and strip without hole (R0).}
\end{figure}

When the hole is located close to the edge of the strip the
maximum of $I_c(B)$-dependence is shifted to the finite magnetic
field (sample X150, black squares in Fig. 7). Nearly linear
$I_c(B)$-dependence at low $B$ again demonstrates that vortices
enter via edges and the shift in $I_c(B)$ is connected with
different vortex entry conditions from the left and right edges
due to presence of the hole. Note, that such a shift was also
observed for strip with central hole (sample R32, red circles in
Fig. 7) where it most probably appears due to relatively large
intrinsic defect at the left edge of the strip (the location of
the defect follows from the direction of the shift in $I_c(B)$) .
\begin{figure}[hbtp]
\includegraphics[width=0.53\textwidth]{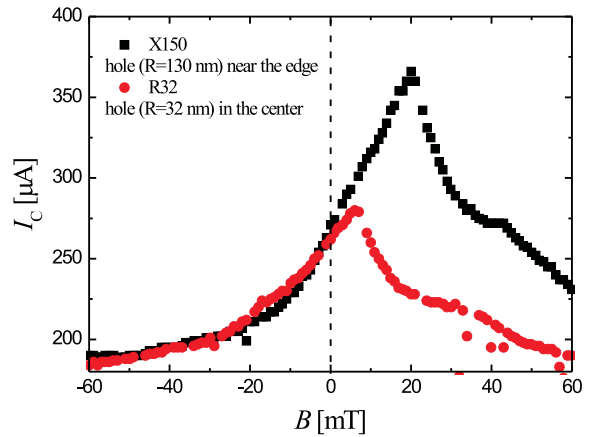}
\caption{Dependence of critical current $I_c$ on magnetic field of
sample X150 (black squares, hole with $R$ = 130 nm is located at
the distance $x$ = 150 nm from left edge; lower left SEM image in
Fig. 2) and sample R32 (red circles; $R$ = 32 nm; the hole is in
the center of strip).}
\end{figure}

\section{Discussion}

There are three main experimental observations which confirm
theoretical predictions formulated in Ref. \cite{SUST_DV} and
\cite{PRB_DV_Z}:

i) the non-monotonic dependence of $I_c$ on coordinate $x$ of hole
across the width of superconducting strip (Fig. 2);

ii) the dependence of $I_c$ on radius of hole which is placed in
the middle of the strip is much weaker than it is expected from
geometrical point of view when radius of hole is in the range $\xi
\ll R \ll W$ (Fig. 5);

iii) the weak dependence of $I_c$ on magnetic field at small $B$
when the hole is located in the middle of strip (Fig. 6). This
weak $I_c(B)$-dependence is caused by generation of
vortex-antivortex (V-A) pairs near the hole due to a current
crowding in vicinity of that place. This area is weakly influenced
by Meissner currents which are mainly concentrated near edges of
strip. Contrary, the strong, linear decrease of $I_c$ with
increasing $B$ is observed when vortex is generated at the edge of
strip. This mechanism dominates over V-A-generation mechanism
mostly in strips with holes which are shifted from the center of
strip or in strips without any artificial holes (Figs. 6 and 7).

Quantitative deviation from the theory we explain by the presence
of the intrinsic defects in real strips. Indeed, some strips
without hole has a critical current comparable to $I_c$ of the
strip with the hole (see Fig. 5). Because measured suppression of
$I_c$ by the circular hole with radius $\xi \ll R \ll W$ is not
large (see Fig. 6) one can imagine situation when intrinsic defect
suppresses $I_c$ stronger than circular hole. This result is
supported by the measurement of $I_c(B)$-dependence which
demonstrates linear variation of $I_c$ at low magnetic field
despite the presence of the hole in the center (for illustration
see Fig. 7, sample R32). We believe, that asymmetry of $I_c(B)$,
visible in Fig. 7 for strip with the central hole also comes from
the intrinsic defects near the hole or deviation from the circular
shape which also breaks mirror symmetry in the superconducting
strip.

Although we made our measurements for relatively wide micron size
strip we believe that obtained results and their analysis are at
least qualitatively valid for narrower strips too. Our numerical
calculations show that the shape of $I_c(x)$-dependence does not
change substantially while one keeps the ratio $R/W$ the same and
when $R \gg \xi$. Our theoretical investigation shows that in the
strips with width $W=20-60 \xi$ the minima of dependence $I_c(x)$
correspond to holes placed on the distance $\sim 2R$ from the edge
of the strip which is close to the experimental result for strip
with $W \simeq 200 \xi$ (see Fig. 2).

From  Fig. 3 it is also seen that $I_c$ is almost independent of
the coordinate in a pretty wide range of $x$ in case the smallest
hole ($R=3\xi$) which is placed in the vicinity of center of the
strip. The plateau in $I_c$(x)-dependence decreases with increase
of $R$ and, in case of the largest radius, a sharp maximum of
$I_c$ is seen in the middle of strip.

Note, that when radius of the hole $R\ll W$ and $R \lesssim 4\xi$
there is relatively strong dependence of $I_c$ on $R$ (see inset
in Fig. 4). It results in smaller value of $I_c$ for the strip
with hole at the edge (in the form of semicircle with radius
$\sqrt{2}R$) than for the strip with central hole (it could be
seen in Fig. 3). For larger radii we have opposite situation (see
Fig. 3 and Fig. 2).

Our results could be applied for understanding of some properties
of superconducting nanowire single-photon detectors (SNSPD). If
one supposes that the absorbed photon creates finite region with
suppressed superconductivity and size smaller then $W$ then it
will influence the critical current of the nanowire in a similar
way like holes studied in this work. Our result directly
demonstrates that the resistive state will appear at different
currents, depending on the place in the strip where the photon was
absorbed. Therefore detection efficiency of SNSPD should change
monotonically with variation of the current, the effect which was
observed in all experiments on SNSPD (see for example recent
review \cite{Natarajan}).

\begin{acknowledgments}

The work was partially supported by the Russian Foundation for
Basic Research (project 15-42-02365/15) and by the Ministry of
education and science of the Russian Federation (the agreement of
August 27, 2013,  N 02.Â.49.21.0003 between The Ministry of
education and science of the Russian Federation and Lobachevsky
State University of Nizhni Novgorod).

\end{acknowledgments}


\end{document}